\documentclass[aps,prb]{revtex4}
\usepackage{amsmath}
\usepackage{graphicx}
\usepackage{amsmath}

\begin{document}

\title{Comment on ``Fully covariant radiation force on a polarizable particle''}
\author{A.I. Volokitin$^{1,2}$\footnote{
Corresponding author. \textit{E-mail address}:alevolokitin@yandex.ru} and
B.N.J. Persson$^2$}
\affiliation{$^1$ Samara State Technical University, 443100 Samara, Russia}
\affiliation{$^2$Peter Gr\"unberg Institute, Forschungszentrum J\"ulich,
D-52425, Germany}

\begin{abstract}
Recently Pieplow and Henkel  (PH) (NJP \textbf{15} (2013) 023027
) presented a new fully covariant theory of the
Casimir friction force acting on small neutral particle moving  parallel to flat surface. We compare results of this theory with results which follow from a fully relativistic theory of friction 
in  plate-plate configurations  in the limit when one plate is considered as sufficiently rarefied.
We show that there is an agreement between these theories.
\end{abstract}

\maketitle

\pagestyle{empty}

\section{Introduction}

All bodies are surrounded by a fluctuating electromagnetic field due to the
thermal and quantum fluctuations of the charge and current density inside
the bodies. Outside the bodies this fluctuating electromagnetic field exists
partly in the form of propagating electromagnetic waves and partly in the
form of evanescent waves. The theory of the fluctuating electromagnetic
field was developed by Rytov \cite{Rytov53,Rytov67,Rytov89}. A great variety
of phenomena such as Casimir-Lifshitz forces \cite{Lifshitz54}, near-field
radiative heat transfer \cite{Polder}, and friction forces \cite
{Volokitin1999,VolokitinRMP2007,VolokitinPRB2008} can be described
using this theory.

 In \cite{Volokitin1999} we used the dynamical
modification of the Lifshitz theory to calculate the friction
force between two plane parallel surfaces in parallel  relative
motion with velocity $V$. The calculation of the van der Waals
friction is more complicated than of the Casimir-Lifshitz force
and the radiative heat transfer because it requires the
determination of the electromagnetic field between moving
boundaries. The solution can be found by writing the boundary
conditions on the surface of each body in the rest reference frame
of this body. The relation between the electromagnetic fields in
the different reference frames is determined by the Lorenz
transformation. In \cite{Volokitin1999} the electromagnetic field in
the vacuum gap between the bodies was calculated to linear order
in $V/c$, which give the contribution to the friction force to
order $(V/c)^2$.  These relativistic corrections were neglected  within the non-relativistic theory developed in\cite{Volokitin1999}. The same non-relativistic theory was used
in\cite {Volokitin01b} to calculate the frictional drag between
quantum wells, and in \cite{Volokitin03a,Volokitin03b} to
calculate the friction force between flat parallel surfaces in
normal relative motion. In Ref. \cite{Volokitin06} we presented a
rigorous quantum mechanical calculation using the Kubo formula
for the friction coefficient. This calculation confirmed the
correctness of the approach based on the dynamical modification of
the Lifshitz theory, at least to linear order in the sliding velocity
$V$. For a review of the van der Waals friction see
\cite{VolokitinRMP2007}.

In Ref. \cite{VolokitinPRB2008} we developed a fully relativistic theory of  the Casimir-Lifshitz forces and
the radiative heat transfer at non-equilibrium conditions, when the
interacting bodies are at different temperatures, and they move relative to
each other with the arbitrary velocity $V$. In comparison with previous
calculations\cite{Volokitin1999,Volokitin01b,Volokitin03a,Volokitin03b}, we
did not make any approximation in the Lorentz transformation of the
electromagnetic field. This allowed us to determine the field in one
 reference frame, knowing the same field in another reference frame.
Thus, the solution of the electromagnetic problem was exact. Knowing the
electromagnetic field we calculated the stress tensor and the Poynting
vector which determined the Casimir-Lifshitz forces and the heat transfer,
respectively. Taking the limit when one of the bodies is rarefied, it is possible to 
obtain  the Casimir-Lifshits force and friction,  and the radiative heat transfer for a small
particle-surface configuration. However, in this approach additional approximations were made which 
did not allow to make detailed comparison with other  theories of friction for the particle-surface configuration in ultra relativistic case.

The problem of friction for a small neutral particle  moving  parallel  to a solid surface (particle-surface configuration) was considered by number of authors (see \cite{VolokitinRMP2007,Dedkov2008,Henkel2013}, and references therein). At present the
interest in this problem is increasing because it is  linked to quantum Cherenkov radiation \cite{Kardar2013}. Recently
a fully covariant theory of friction in particle-surface configuration was proposed by  Pieplow and Henkel  (PH)  \cite{Henkel2013}  and comparison with results of previous authors was given. The theory 
presented by PH  agrees with relativistic theory proposed by Dedkov and Kyasov (DK) \cite{Dedkov2008}. 
However, it is well known that the friction between
a particle and solid surface, mediated by evanescent electromagnetic waves, can be extracted from friction acting between two plates assuming that one plate is sufficiently rarefied \cite{VolokitinRMP2007}. A fully relativistic theory
of friction between two plates in parallel relative motion (plate-plate configuration) was developed in \cite{VolokitinPRB2008}. In the present Comment   the friction in particle-plate configuration
is calculated  from the friction in plate-plate configuration assuming that one plate is sufficiently rarefied. We  compare our results  with the results of Ref. \cite{Henkel2013} and  show that there is  agreement  between  these two theories.

\section{Basic results of a fully relativistic theory of friction between two plates at parallel relative motion}

\begin{figure}[tbp]
\includegraphics[width=0.45\textwidth]{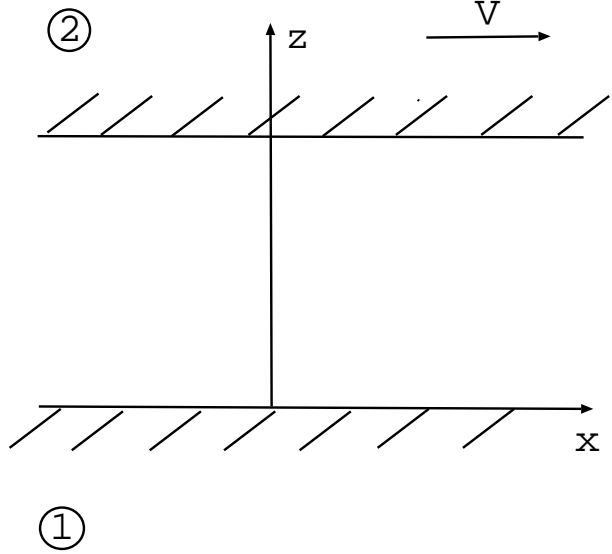}
\caption{Two semi-infinite bodies with plane parallel surfaces separated by
a distance $d$. The upper solids moves parallel to other with velocity $V$. }
\label{Fig1}
\end{figure}

We consider two semi-infinite solids having flat parallel surfaces separated
by a distance $d$ and moving with the velocity $V$ relative to each other,
see Fig. \ref{Fig1}.
We introduce the two coordinate systems $K$ and $K^{\prime }$ with
coordinate axes $xyz$ and $x^{\prime }y^{\prime }z^{\prime }$. In the $K$
system body \textbf{1} is at rest while body \textbf{2} is moving with the
velocity $V$ along the $x-$ axis (the $xy$ and $x^{\prime }y^{\prime }$ planes
are in the surface of body \textbf{1, }$x$ and $x^{\prime}$- axes have the
same direction, and the $z$ and $z^{\prime}-$ axes point toward body \textbf{%
2}). In the $K^{\prime}$ system body \textbf{2} is at rest while body
\textbf{1} is moving with velocity $-V$ along the $x-$ axis.
Since
the system is translational invariant in the $\mathbf{x}=(x,y)$
plane, the electromagnetic field can be represented by the Fourier
integrals
\begin{eqnarray}
\mathbf{E}(\mathbf{x},z,t)=\int_{-\infty}^{\infty} d\omega\int \frac{d^2q}{(2\pi)^2}e^{i\mathbf{q}
\cdot \mathbf{x}-i\omega t}
\mathbf{E}(\mathbf{q},\omega,z),   \\
\mathbf{B}(\mathbf{x},z,t)=\int_{-\infty}^{\infty} d\omega\int \frac{d^2q}{(2\pi)^2}e^{i\mathbf{q}
\cdot \mathbf{x}-i\omega t}
\mathbf{B}(\mathbf{q},\omega,z),
\end{eqnarray}
where $\mathbf{E}$ and $\mathbf{B}$ are the electric and magnetic induction field, respectively,
and $\mathbf{q}$ is the two-dimensional wave vector in $xy$- plane. After
Fourier transformation it is convenient to decompose the electromagnetic field into  $s$- and $p$- polarized components.
For the $p$- and $s$ -polarized electromagnetic waves the electric field $\mathbf{E}(\mathbf{q},\omega,z)$ is in plane of incidence,
and perpendicular to that plane, respectively. In the vacuum
gap between the bodies  the electric field $\mathbf{E}(\mathbf{q},\omega,z)$,  and the  magnetic induction field
$\mathbf{B}(\mathbf{q},\omega,z)$ can be written in the form
\begin{equation}
\mathbf{E}(\mathbf{q},\omega ,z)= \left( v_s\hat{n}_s +
v_p\hat{n}_p^+\right)e^{-k_zz} +\left( w_s\hat{n}_s +
w_p\hat{n}_p^-\right)e^{k_zz}
  \label{one}
\end{equation}
\begin{eqnarray}
\mathbf{B}(\mathbf{q},\omega ,z)=\left(
 v_s\hat{n}_p^{+}  - v_p\hat{n}_s\right)e^{-k_zz}  
 + \left( w_s\hat{n}_p^{-} - w_p\hat{n_s}\right)e^{k_zz} 
 \label{two}
\end{eqnarray}
where 
$k_z=((q^2-(\omega+i0^+/c)^2)^{1/2},\,\hat{n}_s=[\hat{z}\times\hat{q}] = (-q_y, q_x,0)/q,\,\hat{n}_p^{\pm}=[\hat{k}^{\pm}\times
\hat{n}_s]=(\mp q_xik_z,\mp q_yik_z,q^2)/(kq),\, k=\omega/c ,\,\hat{k}^{\pm}=(\mathbf{q}\pm i\hat{z}k_z)/k$.
At the surfaces of the bodies the amplitude
of the outgoing electromagnetic wave must be equal to the amplitude of the reflected
wave plus the amplitude of the radiated wave. Thus, the boundary conditions
for the electromagnetic
field at $z=0$ in the $K$- reference frame
can be written in the form
\begin{equation}
v_{p(s)}=R_{1p(s)}(\omega,q )w_{p(s)}+E^{f}_{1p(s)}(\omega,q)
\label{three}
\end{equation}
where $R_{1p(s)}(\omega )$ is the reflection amplitude for surface
\textbf{1} for the $p(s)$ - polarized electromagnetic field, and
where $ E^{f}_{1p(s)}(\omega )$ is the amplitude of the
 fluctuating electric field radiated  by body \textbf{1} for a $p(s)$-polarized wave.

In the $K^{^{\prime }}$- reference frame  the electric field can be written in the form
\begin{equation}
\mathbf{E}^{\prime}(\mathbf{q}^{\prime},\omega^{\prime} ,z)=
\left( v_s^{\prime}\hat{n}_s^{\prime} +
v_p^{\prime}\hat{n}_p^{\prime +}\right)e^{-k_zz} +\left(
w_s^{\prime}\hat{n}_s^{\prime} + w_p^{\prime}\hat{n}_p^{\prime
-}\right)e^{k_zz}
  \label{four}
\end{equation}
where $\mathbf{q}^{\prime}=(q_x^{\prime},q_y,0),\,q_x^{\prime}=(q_x - \beta k)\gamma,\,\omega^{\prime}=
(\omega - Vq_x)\gamma,\,
\gamma=1/\sqrt{1-\beta^2},\,\beta = V/c,\, \hat{n}^{\prime}_s =
(-q_y, q^{\prime}_x,0)/q^{\prime},\,\hat{n}^{\prime \pm}_{p}=
(\mp q^{\prime}_xk_z,\mp q_yk_z,q^{\prime 2})/(k^{\prime}q^{\prime}),$
\[
q^{\prime}=\gamma\sqrt{q^2-2\beta kq_x+\beta^2(k^2-q_y^2)}.
\]
The boundary conditions at $z=d$ in the $K^{^{\prime }}$- reference frame can be written in a form similar to
Eq. (\ref{three}):
\begin{equation}
w^{\prime}_{p(s)}=e^{-2k_zd}R_{2p(s)}(\omega^{\prime},q^{\prime} )v^{\prime}_{p(s)}
+e^{-k_zd}E^{\prime f}_{2p(s)}(\omega^{\prime},q^{\prime} ),
\label{five}
\end{equation}
where $R_{2p(s)}(\omega )$ is the reflection amplitude for surface
\textbf{2} for $p(s)$ - polarized electromagnetic field, and where
$ E^{f}_{2p(s)}(\omega )$ is the amplitude of the
 fluctuating electric field radiated by  body \textbf{2} for a $p(s)$-polarized wave.
A Lorentz transformation for the electric field gives
\begin{equation}
E_x^{\prime}=E_x,\,E_y^{\prime} = (E_y-\beta B_z)\gamma,\, E_z^{\prime} = (E_z+\beta B_y)\gamma
\label{six}
\end{equation}

Using Eqs. (\ref{one},\ref{two},\ref{four}) and (\ref{six}) we get
\begin{equation}
v_p^{\prime} = \frac{k^{\prime}\gamma}{kqq^{\prime}}\left[-i\beta k_zq_yv_s + (q^2 - \beta kq_x)v_p\right],
\label{seven}
\end{equation}
\begin{equation}
w_p^{\prime} = \frac{k^{\prime}\gamma}{kqq^{\prime}}\left[i\beta k_zq_yw_s + (q^2 - \beta kq_x)w_p\right],
\label{eight}
\end{equation}
\begin{equation}
v_s^{\prime} = \frac{k^{\prime}\gamma}{kqq^{\prime}}\left[i\beta k_zq_yv_p + (q^2 - \beta kq_x)v_s\right],
\label{nine}
\end{equation}
\begin{equation}
w_s^{\prime} = \frac{k^{\prime}\gamma}{kqq^{\prime}}\left[-i\beta k_zq_yw_p + (q^2 - \beta kq_x)w_s\right].
\label{ten}
\end{equation}

Substituting Eqs. (\ref{seven}-\ref{ten}) in Eq. (\ref{five}) and using Eq. (\ref{three}) we get
\[
(q^2 - \beta kq_x)\Delta_{pp}w_p+i\beta k_zq_y\Delta_{sp}w_s
\]
\begin{equation}
=e^{-2k_zd}R_{2p}^{\prime}\left[(q^2 - \beta kq_x)E_{1p}^f-i\beta k_zq_yE_{1s}^f\right] +
\frac{kqq^{\prime}}{k^{\prime}\gamma}e^{-k_zd}E_{2p}^{\prime f},
\label{eleven}
\end{equation}
\[
(q^2 - \beta kq_x)\Delta_{ss}w_s - i\beta k_zq_y\Delta_{ps}w_p
\]
\begin{equation}
=e^{-2k_zd}R_{2s}^{\prime}\left[(q^2 - \beta kq_x)E_{1s}^f + i\beta k_zq_yE_{1p}^f\right] +
\frac{kqq^{\prime}}{k^{\prime}\gamma}
e^{-k_zd} E_{2s}^{\prime f},
\label{twelve}
\end{equation}
where
\[
\Delta_{pp} = 1 - e^{-2k_zd}R_{1p}R_{2p}^{\prime},\,  \Delta_{ps} = 1 + e^{-2k_zd}R_{1p}R_{2s}^{\prime},
\]
$\Delta_{ss}=\Delta_{pp}(p\leftrightarrow s)$, $\Delta_{sp}=\Delta_{ps}(p\leftrightarrow s)$, $R_{2p(s)}^{\prime}=R_{2p(s)}(\omega^{\prime},q^{\prime})$, the symbol ($p\leftrightarrow s$) means permutation of the
indexes $p$ and $s$.
From Eqs. (\ref{eleven},\ref{twelve})  and (\ref{three}) we get

\[
w_p = \Big\{\left[(q^2 - \beta kq_x)^2R_{2p}^{\prime}\Delta_{ss} + \beta^2 k_z^2q_y^2R_{2s}^{\prime}\Delta_{sp}\right]
E_{1p}^fe^{-2k_zd}
\]
\[
-i\beta k_zq_y(q^2 - \beta kq_x)(R_{2p}^{\prime} +
R_{2s}^{\prime})E_{1s}^fe^{-2k_zd}
\]
\begin{equation}
+ \frac{kqq^{\prime}}{k^{\prime}\gamma}\left[(q^2 - \beta
kq_x)\Delta_{ss}E_{2p}^{\prime f}- i\beta k_zq_y\Delta_{sp}E_{2s}^{\prime
f}\right]e^{-k_zd} \Big\}\Delta^{-1}, \label{thirteen}
\end{equation}
\[
v_p = \Big\{\left[(q^2 - \beta kq_x)^2\Delta_{ss} - \beta^2 k_z^2q_y^2\Delta_{sp}\right]
E_{1p}^f
\]
\[
-i\beta k_zq_y(q^2 - \beta kq_x)R_{1p}(R_{2p}^{\prime} + R_{2s}^{\prime})e^{-2k_zd}E_{1s}^f
\]
\begin{equation}
+ \frac{kqq^{\prime}}{k^{\prime}\gamma}R_{1p}\left[(q^2 - \beta kq_x)\Delta_{ss}E_{2p}^{\prime f}-
i\beta k_zq_yD_{sp}E_{2s}^{\prime f}\right]e^{-k_zd}
\Big\}\Delta^{-1},
\label{fourteen}
\end{equation}

\[
w_s = \Big\{\left[(q^2 - \beta kq_x)^2R_{2s}^{\prime}\Delta_{pp} + \beta^2 k_z^2q_y^2R_{2p}^{\prime}\Delta_{ps}\right]
E_{1s}^fe^{-2k_zd}
\]
\[
+i\beta k_zq_y(q^2 - \beta kq_x)(R_{2p}^{\prime} + R_{2s}^{\prime})E_{1p}^fe^{2ik_zd}
\]
\begin{equation}
+ \frac{kqq^{\prime}}{k^{\prime}\gamma}\left[(q^2 - \beta kq_x)D_{pp}E_{2s}^{\prime f}+
i\beta k_zq_yD_{ps}E_{2p}^{\prime f}\right]e^{-k_zd}
\Big\}\Delta^{-1},
\label{fifteen}
\end{equation}
\[
v_s = \Big\{\left[(q^2 - \beta kq_x)^2\Delta_{pp} - \beta^2 k_z^2q_y^2\Delta_{ps}\right]
E_{1s}^f
\]
\[
+i\beta k_zq_y(q^2 - \beta kq_x)R_{1p}(R_{2p}^{\prime} + R_{2s}^{\prime})e^{-2k_zd}E_{1p}^f
\]
\begin{equation}
+ \frac{kqq^{\prime}}{k^{\prime}\gamma}R_{1s}\left[(q^2 - \beta
kq_x)\Delta_{pp}E_{2s}^{\prime f}+ i\beta k_zq_y\Delta_{ps}E_{2p}^{\prime
f}\right]e^{-k_zd} \Big\}\Delta^{-1}, \label{sixteen}
\end{equation}
where
\[
\Delta = (q^2 - \beta kq_x)^2\Delta_{ss}\Delta_{pp} - \beta^2k_z^2q_y^2\Delta_{ps}\Delta_{sp}.
\]

The fundamental characteristic
of the fluctuating electromagnetic field is the correlation function, determining
the average product of amplitudes $E^{f}_{p(s)}(\mathbf{q},\omega)$.
According to the
general theory of the fluctuating electromagnetic field
(see for a example \cite{VolokitinRMP2007}):
\[
<|E^{f}_{p(s)}(\mathbf{q},\omega)|^2>=\frac{\hbar \omega^2i}
{2c^2|k_z|^2} \left(n(\omega)+\frac{1}{2}\right)
[(k_z-k_z^*)(1-|R_{p(s)}|^2)
\]
\begin{equation}
+(k_z+k_z^*)(R_{p(s)}^*-R_{p(s)})]  \label{seventeen}
\end{equation}
where $<...>$ denote statistical average over the random field.
We note that $k_z$ is  purely imaginary ($k_z=-i|k_z|$) for $q<\omega/c$ (propagating waves), and real
 for  $q>\omega/c$ (evanescent waves). The  Bose-Einstein factor
\[
n(\omega )=\frac 1{e^{\hbar \omega /k_BT}-1}.
\]
Thus for $q<\omega/c$
and $q>\omega/c$ the
correlation functions are determined by the first and the second
terms in Eq. (\ref{seventeen}), respectively.

The force which acts  on the surface of body \textbf{1} can be calculated from the Maxwell stress tensor
$\sigma_{ij}$, evaluated at $z=0$:
\[
\sigma _{ij} =\frac 1{4\pi }\int_0^\infty d\omega \int \frac{d^2q}{(2\pi)^2}
\Big[ <E_iE_j^*> + <E_i^*E_j> + <B_iB_j^*> + <B_i^*B_j>
\]
\begin{equation}
 - \delta_{ij}(<\mathbf{E\cdot E^*}> + <\mathbf{B\cdot B^*}>)\Big] _{z=0}   \label{3one}
\end{equation}
Using Eqs. (\ref{one},\ref{two}) for the $x$ - component of the force we get
\[
\sigma_{xz} =\frac i{4\pi }\int_0^{\infty }d\omega \int \frac{d^2q}{(2\pi)^2}\frac{q_x}{k^2} \left[
(k_z-k_z^{*})\left (\left\langle \mid w_p\mid
^2\right\rangle +\left\langle \mid w_s\mid
^2\right\rangle
\right. \right.
\]
\begin{equation}
\left.\left. -\left\langle \mid v_p\mid ^2\right\rangle -
\left\langle \mid v_s\mid ^2\right\rangle \right)
+(k_z+k_z^{*})\left\langle w_pv_p^* + w_sv_s^*
 - c.c\right\rangle \right]
\label{2two}
\end{equation}
Substituting Eqs.  (\ref{thirteen}-\ref{sixteen}) for the amplitudes of the
electromagnetic field in Eq. (\ref{2two}),  and performing averaging over the fluctuating electromagnetic field with the help of
Eq. (\ref{seventeen}), we get the $x$-component of the force \cite{VolokitinPRB2008}
\[
F_x = \sigma_{xz} =\frac \hbar {8\pi ^3}\int_0^\infty d\omega \int_{q<\omega /c}d^2q\frac{q_x}{|\Delta|^2}[(q^2 - \beta kq_x)^2 - \beta^2k_z^2q_y^2]
\]
\[
\times  [(q^2 - \beta kq_x)^2(1-\mid R_{1p}\mid ^2)(1-\mid R_{2p}^{\prime }\mid ^2)|\Delta_{ss}|^2
\]
\[
-\beta^2k_z^2q_y^2(1-\mid R_{1p}\mid ^2)(1-\mid R_{2s}^{\prime }\mid ^2)|\Delta_{sp}|^2 + (p\leftrightarrow s)]
\left( n_2(\omega^{\prime})-n_1(\omega )\right)
\]
\[
+\frac \hbar {2\pi ^3}\int_0^\infty d\omega \int_{q>\omega
/c}d^2q\frac{q_x}{|\Delta|^2}[(q^2 - \beta kq_x)^2 - \beta^2k_z^2q_y^2]
e^{-2 k_z d}
\]
\[
\times [(q^2 - \beta kq_x)^2 \mathrm{Im}R_{1p}\mathrm{Im}R_{2p}^{\prime}|\Delta_{ss}|^2+ \beta^2k_z^2q_y^2
\mathrm{Im}R_{1p}\mathrm{Im}
R_{2s}^{\prime}|\Delta_{sp}|^2
\]
\begin{equation}
+ (p\leftrightarrow s)]\left( n_2(\omega^{\prime})-n_1(\omega
)\right).   \label{2three}
\end{equation}
 The symbol $(p\leftrightarrow s)$
denotes the terms which can be obtained from the preceding terms
by permutation of the indexes $p$ and $s$. The first term in Eq.
(\ref{2three}) represents the contribution to the friction from
propagating waves ($q<\omega /c$), and the second term  from the
evanescent waves ($q>\omega /c$).

\section{A fully relativistic theory of the Casimir force and friction force, and radiated heat transfer for a small particle moving parallel to a flat surface}
According to Eq. (\ref{2three}) the contribution to the friction force from the evanescent waves is given by
\[
F_x=\frac \hbar {2\pi ^3}\int_0^\infty d\omega \int_{q>\omega
/c}d^2q\frac{q_x}{|\Delta|^2}[(q^2 - \beta kq_x)^2 - \beta^2k_z^2q_y^2]
e^{-2 k_z d}
\]
\begin{equation}
\times [ \mathrm{Im}R_{1p}\mathrm{Im}\Delta_p +(p\leftrightarrow s)]\left( n_2(\omega^{\prime})-n_1(\omega)\right),   \label{2three}
\end{equation}
where
\[
\Delta_p=(q^2 - \beta kq_x)^2R_{2p}^{\prime}|\Delta_{ss}|^2+ \beta^2k_z^2q_y^2
R_{2s}^{\prime}|\Delta_{sp}|^2,
\]
$\Delta_s=\Delta_p(p\leftrightarrow s)$. 
If in Eq. (\ref{2three}) one
neglects the terms of the order $\beta^2$ then the contributions from waves
with $p$- and $s$- polarization will be separated. In this case Eq. (\ref%
{2three}) is reduced to the formula obtained in \cite{Volokitin1999}
\begin{equation}
F_x =
\frac \hbar {2\pi ^3}\int_0^\infty d\omega \int_{q>\omega /c}d^2q q_xe^{-2 k_z d}
\left(\frac{\mathrm{Im}R_{1p}\mathrm{Im}R_{2p}^{\prime}}{|\Delta_{pp}|^2} + \frac{\mathrm{Im}R_{1s}\mathrm{Im}R_{2s}^{\prime}}{|\Delta_{ss}|^2}\right)\left( n_2(\omega^{\prime})-n_1(\omega )\right),
\label{Vol1998}
\end{equation}
Thus, to
the order $\beta^2$ the mixing of waves with different polarization can be
neglected, what agrees with the results obtained in \cite{Volokitin1999}. At $
T=0$ K the propagating waves do not contribute to friction but the
contribution from evanescent waves is not equal to zero. Taking into account
that $n(-\omega)= -1 - n(\omega)$ from Eq. (\ref{2three}) we get the
friction mediated by the evanescent electromagnetic waves at zero
temperature (in literature this type of friction is denoted as quantum
friction \cite{Pendry1997})
\begin{equation*}
F_x = - \frac \hbar {\pi ^3}\int_0^\infty dq_y \int_0^\infty dq_x
\int_0^{q_xV} d\omega \frac{q_x}{|\Delta|^2}[(q^2 - \beta kq_x)^2 -
\beta^2k_z^2q_y^2] e^{-2 k_z d}
\end{equation*}
\begin{equation}
\times [\mathrm{Im}R_{1p}\mathrm{Im}\Delta_p + \mathrm{Im}R_{1s}\mathrm{Im}\Delta_s].
\label{zerotemfr}
\end{equation}

If in Eq. (\ref{zerotemfr}) one
neglects the terms of the order $\beta^2$ then the contributions from waves
with $p$- and $s$- polarization will be separated. In this case Eq. (\ref%
{zerotemfr}) is reduced to the formula obtained by Pendry for $p$-polarized  waves in the non-retarded limit \cite{Pendry1997}
\begin{equation}
F_x =
-\frac \hbar {\pi ^3}\int_0^\infty dq_y \int_0^\infty dq_x
\int_0^{q_xV} d\omega q_x
\left(\frac{\mathrm{Im}R_{1p}\mathrm{Im}R_{2p}^{\prime}}{|\Delta_{pp}|^2} + \frac{\mathrm{Im}R_{1s}\mathrm{Im}R_{2s}^{\prime}}{|\Delta_{ss}|^2}\right)e^{-2 k_z d},
\label{Pendry1997}
\end{equation}

The  friction force acting on a small particle moving in  parallel
to a flat surface can be obtained from the friction between two
semi-infinite bodies in the limit when one of the bodies is
sufficiently rarefied.  We will assume that the rarefied body consists of
small  particles which  have   electric dipole moments. We assume that the dielectric permittivity  of this body, say body \textbf{2}, is close to the
unity, i.e. $\varepsilon _2-1\rightarrow 4\pi n\alpha\ll 1$ , where $n$ is the
concentration of particles in body \textbf{2} in the co-moving reference frame $K^{\prime}$,  $\alpha $ is  their electric polarizability. To
linear order in the concentration $n$ the reflection amplitudes
are
\[
R^{\prime}_{2p} = \frac{\varepsilon _2^{\prime}k_z - \sqrt{k_z^2 - (\varepsilon _2^{\prime}-1)k^{\prime 2}}} {\varepsilon _2^{\prime}k_z +
\sqrt{k_z^2 - (\varepsilon ^{\prime}_2-1) k^{\prime 2}}}
\approx \frac{\varepsilon ^{\prime} _2-1}{4}\frac{q^{\prime 2}+k_z^2}{k_z^2} = n\pi\frac{q^{\prime 2}+k_z^2}{k_z^2} \alpha^{\prime},
\]
\[
R_{2s} = \frac{k_z - \sqrt{k_z^2 - (\varepsilon _2^{\prime}-1) k^2}}{k_z +\sqrt{k_z^2 - (\varepsilon _2^{\prime}-1) k^2}}
\approx \frac{\varepsilon^{\prime} _2-1}{4}\frac{q^{\prime 2}-k_z^2}{k_z^2} = n\pi\frac{q^{\prime 2}-k_z^2}{k_z^2} \alpha^{\prime}.
\]
To
linear order in the concentration $n$ the functions $\Delta_{pp},\,\Delta_{ss},\,\Delta_{sp}$ and $\Delta_{ps}$ should be calculated at $n=0$. Using that $\Delta_{pp}=\Delta_{ss}=\Delta_{sp}=\Delta_{ps}
=1$ for $n=0$, we get 
\[
\Delta = (q^2 - \beta kq_x)^2 - \beta^2k_z^2q_y^2= \frac{(qq^{\prime})^2}{\gamma^2},
\]
\[
\Delta_p=\{q^{\prime 2}[(q^2 - \beta kq_x)^2 +
\beta^2k_z^2q_y^2] +k_z^2[(q^2 - \beta kq_x)^2 -
\beta^2k_z^2q_y^2]\} \frac{\pi n \alpha^{\prime}}{k_z^2}
\]
\[
=q^{\prime 2}\{q^2[k_z^2 + (k-\beta q_x)^2] +k_z^2[q^2-2\beta^2q_x^2]\}\frac{\pi n \alpha^{\prime}}{k_z^2}
\]
\[
=q^{\prime 2}\{q^2(k-\beta q_x)^2 +2k_z^2(q^2-\beta^2q_x^2)\}\frac{\pi n \alpha^{\prime}}{k_z^2},
\]
\[
\Delta_s=\{q^{\prime 2}[(q^2 - \beta kq_x)^2 +
\beta^2k_z^2q_y^2] -k_z^2[(q^2 - \beta kq_x)^2 -
\beta^2k_z^2q_y^2]\} \frac{\pi n \alpha^{\prime}}{k_z^2}
\]
\[
=q^{\prime 2}\{q^2[k_z^2 + (k-\beta q_x)^2] -k_z^2[q^2-2\beta^2q_y^2]\}\frac{\pi n \alpha^{\prime}}{k_z^2}
\]
\[
=q^{\prime 2}\{q^2(k-\beta q_x)^2 +2k_z^2\beta^2q_y^2\}\frac{\pi n \alpha^{\prime}}{k_z^2},
\]
where $\alpha^{\prime}=\alpha(\omega^{\prime})$.

The friction force acting on a particle moving  parallel to a
plane surface can be obtained as the ratio between the change of
the frictional shear stress between two surfaces after
displacement of body \textbf{2} by small distance $dz$, and the
number of the particles in a slab with thickness $dz$:

\begin{equation}
f_x^{part}=\frac{dF_x(z)}{n^{\prime}dz}\Big |_{z=d}=
\frac \hbar {\gamma\pi ^2}\int_0^\infty d\omega \int_{q>\omega /c}d^2q\frac{q_x}{
k_z} e^{-2 k_z d} [\mathrm{Im}R_{1p}(\omega)\phi_p + \mathrm{Im}R_{1s}(\omega)\phi_s]\mathrm{Im}\alpha(\omega^{\prime})\left( n_2(\omega^{\prime})-n_1(\omega )\right),
\label{frparticle}
\end{equation}

where $n^{\prime}=\gamma n$  is the
concentration of particles in body \textbf{2} in the  reference frame $K$
\[
\phi_p=(\omega^{\prime}/c)^2+2\gamma^2(q^2-\beta^2q_x^2)\frac{k_z^2}{q^2}
\]
\[
\phi_s=(\omega{^\prime}/c)^2+2\gamma^2\beta^2q_y^2\frac{k_z^2}{q^2}
\]
At $T_2=T_1=0$ K we get
\begin{equation}
f_x^{part}=
-\frac \hbar {\gamma\pi ^2} \int_{-\infty}^{\infty}dq_y\int_{0}^{\infty}dq_x\int_0^{q_xV} d\omega\frac{q_x}{
k_z} e^{-2 k_z d}[\mathrm{Im}R_{1p}(\omega)\phi_p + \mathrm{Im}R_{1s}(\omega)\phi_s]\mathrm{Im}\alpha(\omega^{\prime})
\label{zerotemperaturepart}
\end{equation}
For $\beta^2\ll 1$ and $q\gg \omega/c$, Eq.(\ref{frparticle}) is reduced to the result of non-relativistic theory \cite{VolokitinRMP2007}
\begin{equation}
f_x^{part}=
\frac {2\hbar}  {\pi ^2}\int_0^\infty d\omega \int_{q>\omega /c}d^2q q_x
 qe^{-2 q d}\mathrm{Im}R_{1p}\mathrm{Im}\alpha(\omega-q_xv))\left( n_2(\omega^{\prime})-n_1(\omega )\right),
\label{frparticlenonrel}
\end{equation}

The heat absorbed by the body \textbf{1} in the $K$ system in the plate-plate configuration is determined by the expression which is very similar to the expression for the friction force (\ref{2three}) \cite{VolokitinPRB2008}
\begin{equation}
P_1 =
\frac \hbar {2\pi ^3}\int_0^\infty d\omega \int_{q>\omega /c}d^2q\frac{\omega}{%
|\Delta|^2}[(q^2 - \beta kq_x)^2 - \beta^2k_z^2q_y^2] e^{-2 k_z d}
[\mathrm{Im}R_{1p}\mathrm{Im}\Delta_p + \mathrm{Im}R_{1s}\mathrm{Im}\Delta_s]\left( n_2(\omega^{\prime})-n_1(\omega )\right),
\label{2threepar}
\end{equation}

Using result obtained for the friction in the particle-surface configuration from (\ref{2threepar}) and (\ref{frparticle}) we get the heat absorbed by plate in the $K$ system in the particle-plate configuration
\begin{equation}
P_1^{part}=
\frac \hbar {\gamma\pi ^2}\int_0^\infty d\omega \int_{q>\omega /c}d^2q\frac{\omega}{
k_z} e^{-2 k_z d} [\mathrm{Im}R_{1p}(\omega)\phi_p + \mathrm{Im}R_{1s}(\omega)\phi_s]\mathrm{Im}\alpha(\omega^{\prime})\left( n_2(\omega^{\prime})-n_1(\omega )\right),
\label{heatparticle}
\end{equation}
The heat absorbed by a particle  in the $K^{\prime}$ system ($P_2^{\prime}$) can be obtained from the relation
\begin{equation}
f_xV=P_1+\frac{P_2^{\prime}}{\gamma},
\label{relation}
\end{equation}
which follows from the Lorentz transformation of the Poynting vector. From (\ref{relation}) we get
\begin{equation}
P_2^{\prime}=
\frac \hbar {\gamma\pi ^2}\int_0^\infty d\omega \int_{q>\omega /c}d^2q\frac{\omega}{
k_z} e^{-2 k_z d} [\mathrm{Im}R_{1p}(\omega^{\prime})\phi_p + \mathrm{Im}R_{1s}(\omega^{\prime})\phi_s]\mathrm{Im}\alpha(\omega)\left( n_1(\omega^{\prime})-n_2(\omega )\right),
\label{heatparticle}
\end{equation}
where we transformed variables $\omega,q_x$ in the integrands (\ref{frparticle}) and (\ref{heatparticle}) to $\omega^{\prime},q_x^{\prime}$ using the fact that the Lorentz transformation has unit Jacobian. After such changing we denoted``dummi'' variable $\omega^{\prime},q_x^{\prime}$ as $\omega,\,q_x$. 

The Casimir force  between two moving plates mediated by the evanescent waves is given by \cite{VolokitinPRB2008}
\[
F_z  =\frac \hbar {4\pi ^3}\mathrm{Im}\int_0^\infty d\omega \int_{q>\omega
/c} d^2q\frac{k_z}{\Delta}e^{-2k_zd}
[R_{1p}\Delta_{1p}+R_{1s}\Delta_{1s}]
[1 + n_1(\omega) +
n_2(\omega^{\prime})]
\]
\[
+\frac \hbar {4\pi ^3}\int_0^\infty d\omega \int_{q>\omega
/c}d^2q\frac{k_z}{|\Delta|^2}[(q^2 - \beta kq_x)^2 -
\beta^2k_z^2q_y^2] e^{-2 k_z d}
\]
\begin{equation}
\times \{\mathrm{Im}R_{1p}\mathrm{Re}\Delta_{p}-\mathrm{Re}R_{1p}\mathrm{Im}\Delta_{p}
+ (p\leftrightarrow s)\}\left( n_1(\omega
)-n_2(\omega^{\prime})\right). \label{2five}
\end{equation}
where 
\[
\Delta_{1p}=(q^2 -\beta kq_x)^2R_{2p}^{\prime}\Delta_{ss}+\beta^2k_z^2q_y^2R_{2s}^{\prime}\Delta_{sp},
\]
$\Delta_{1s}=\Delta_{1p}(p\leftrightarrow s)$. In the limit $n\rightarrow 0$: $\Delta_{1p(s)}=\Delta_{p(s)}$.
After similar calculations  as above for the Casimir force acting on a small particle moving parallel to a flat surface we get
\[
F_z^{part}  =\frac \hbar {2\gamma \pi ^2}\int_0^\infty d\omega \int_{q>\omega
/c} d^2qe^{-2k_zd}\Bigg\{
[\phi_p\mathrm{Im}R_{1p}+\phi_s\mathrm{Im}R_{1s}]\mathrm{Re}\alpha^{\prime}\mathrm{coth}\left(\frac{\hbar \omega}{k_BT_1}\right) 
\]
\begin{equation}+
[\phi_p\mathrm{Re}R_{1p}+\phi_s\mathrm{Re}R_{1s}]\mathrm{Im}\alpha^{\prime}\mathrm{coth}\left(\frac{\hbar \omega^{\prime}}{k_BT_2}\right)\Bigg\}
\end{equation}

\section{Comparison with the previous results }

Recently Pieplow and Henkel  \cite{Henkel2013}  presented a fully covariant theory of the
Casimir force and friction force acting on small neutral particle moving parallel to flat surface.
This theory is in  agreement with relativistic theory presented by Dedkov and Kyasov \cite{Dedkov2008}.  In this Comment we have shown that the results of PH and DK for contribution to friction from evanescent waves in the particle-plate configuration are determined by the first derivative of friction force in the plate-plate configuration assuming that one of the plate is sufficiently rarefied. However, inverse procedure is not possible. It is not possible to recover the whole function knowing only its first derivative. The contribution to friction from the propagating waves is more delicate. To make comparison between contributions to friction from propagating waves  in the the particle-plate configuration and the plate-plate configuration it is necessary to consider slab with finite thickness and calculate the friction force acting on the both side of the slab. In contrast to the evanescent waves, which do not contribute to the friction force acting on the back side of the slab,  the propagating waves contribute to the friction force acting on the both side of the slab. However, for large velocities (for example, above the Cherenkov threshold velocity) the friction is dominated by quantum friction determined by the evanescent waves.    

A.I.V acknowledges financial support from the Russian Foundation for Basic
Research (Grant N 14-02-00297-a) and COST Action MP1303 "Understanding and Controlling Nano and Mesoscale Friction.

\end{document}